\documentclass[12pt,fleqn,twoside]{article}%
\usepackage{amsfonts}
\usepackage{espcrc1}
\usepackage{amsmath}
\usepackage{amssymb}
\usepackage{graphicx}%
\providecommand{\U}[1]{\protect\rule{.1in}{.1in}}
\newcommand{\ttbs}{\char'134}

\begin{document}

\title{\textbf{Determination of the propagator in QED}$_{3}$\textbf{ by spectral
function}}
\author{Yuichi Hoshino\\Kushiro National College of Technology,Kushiro City,Hokkaido 084-0916,Japan }
\maketitle

\begin{abstract}
To study the infrared behaviour of the propagator, exponentiation of the
lowest order spectral function has been known.We show this method is helpful
in super renormalizable theory with dimension-full coupling constant.In the
$1/N$ approximation anomalous dimension is independent of $N$,which plays an
important role for confinement and pair condensation.

\end{abstract}

%

\title
{Elsevier instructions for the preparation of a camera-ready paper in \LaTeX}%
%

\author{P. de Groot\address[MCSD]{Mathematics and Computer Science Division,
Elsevier Science Publishers B.V., \\
P.O. Box 103, 1000 AC Amsterdam, The Netherlands}        \thanks
{Footnotes should appear on the first page only to
indicate your present address (if different from your
normal address), research grant, sponsoring agency, etc.
These are obtained with the {\tt\ttbs thanks} command.},
R. de Maas\addressmark\thanks{For following authors with the same
address use the {\tt\ttbs addressmark} command.},
X.-Y. Wang\address{Economics Department, University of Winchester, \\
2 Finch Road, Winchester, Hampshire P3L T19, United Kingdom}
and
A. Sheffield\addressmark[MCSD]\thanks{To reuse an addressmark
later on, label the address with an optional argument to the
{\tt\ttbs address} command, e.g. {\tt\ttbs address[MCSD]}%
, and repeat the label
as the optional argument to the {\tt\ttbs addressmark}
command, e.g. {\tt\ttbs addressmark[MCSD]}.}}%
%

\maketitle
%

\begin{abstract}%
\end{abstract}%

\section{INTRODUCTION}

It has long been known that the infrared behavour of the propagator in the
theory with massless particle as photon is determined by low-energy theorem or
renormalization group analysis[1,2,3].The answer is known as%

\begin{equation}
S_{F}(p)\simeq\frac{Z_{2}(\gamma\cdot p+m)}{(p^{2}-m^{2})^{1-D}}%
,D=\frac{\alpha(d-3)}{2\pi},\alpha=\frac{e^{2}}{4\pi},
\end{equation}

near $p^{2}=m^{2},$where $d$ is a covariant gauge fixing parameter and
$\alpha$ is a fine structure constant.$D$ is an anomalous dimension of wave
function renormalization constant.This formula is valid in the infrared
region.But the super renormalizable model as QED$_{3}$ suggests the validity
of the above type of formula in all region of momentum.If we sum all ladder
type diagrams which is dominant in the infrared we obtain the similar form of
the propagator with dimensionless coupling constant $D=e^{2}/m.$Including
vacuum polarization the infrared divergences turns out to be soft.

\section{\bigskip DIRECT EVALUATION OF THE SPECTRAL FUNCTION}

The spectral representation of the propagator is given[1]%

\begin{align}
S_{F}(s^{\prime})  &  =P\int ds\frac{\gamma\cdot p\rho_{1}(s)+\rho_{2}%
(s)}{s^{\prime}-s}+i\pi(\gamma\cdot p\rho_{1}(s^{\prime})+\rho_{2}(s^{\prime
})),\\
\rho(p)  &  =\frac{1}{\pi}\operatorname{Im}S_{F}(p)=\gamma\cdot p\rho
_{1}(p)+\rho_{2}(p)\nonumber\\
&  =(2\pi)^{2}\sum_{n}\delta(p-p_{n})\int d^{3}x\exp(-ip\cdot x)\left\langle
0|\psi(x)|n\right\rangle \left\langle n|\overline{\psi}(0)|0\right\rangle .
\end{align}

If the intermediate states contain massless particle as photon there are
infrared divergences.In the quenched approximation the state $|n>$ stands for
a fermion and arbitrary numbers of photons,%
\begin{equation}
|n>=|r;k_{1},...,k_{n}>,r^{2}=m^{2},
\end{equation}
we have the solution for the spectral function $\rho(p)$ which is written
symbolically%
\begin{align}
\rho(p)  &  =\int d^{3}x\exp(-ip\cdot x)\int\frac{md^{2}r}{r^{0}}\sum
_{n=0}^{\infty}\frac{1}{n!}\left(  \int\frac{d^{3}k}{(2\pi)^{2}}\theta
(k^{0})\delta(k^{2})\sum_{\epsilon}\right)  _{n}\nonumber\\
&  \times\delta(p-r-\sum_{i=1}^{n}k_{i})\left\langle \Omega|\psi
(x)|r;k_{1},...,k_{n}\right\rangle \left\langle r;k_{1},...,k_{n}%
|\overline{\psi}(0)|\Omega\right\rangle .
\end{align}
Here the notations
\begin{equation}
(f(k))_{0}=1,(f(k))_{n}=%
{\displaystyle\prod\limits_{i=1}^{n}}
f(k_{i})
\end{equation}
have been introduced to show the phase space of each photons.To evaluate the
contribution of soft photons,first we consider the situation when only the
$n$-th photon is soft.We define the matrix element
\begin{equation}
T_{n}=\left\langle \Omega|\psi|r;k_{1},....,k_{n}\right\rangle .
\end{equation}
We consider $T_{n}$ for $k_{n}^{2}\neq0,$continue off the photon mass shell
\begin{align}
T_{n}  &  =\epsilon_{n}^{\mu}T_{n\mu},\nonumber\\
T_{n\mu}  &  =-\frac{i}{\sqrt{Z_{3}}}\int d^{3}y\exp(ik_{n}\cdot
y)\left\langle \Omega|T\psi(x)j_{\mu}(y)|r;k_{1},...,k_{n-1}\right\rangle .
\end{align}
$T_{n}$ satisfies Ward-Takahashi-identity%
\begin{equation}
k_{n\mu}T_{\mu}^{n}(r;k_{1},...k_{n})=eT_{n-1}(r;k_{1},...k_{n-1}),r^{2}%
=m^{2}.
\end{equation}
Formal proof of Ward-Takahahsi-identity is provided with LSZ reduction formula
and subsidary condition for photon[4]
\begin{equation}
\partial\cdot A^{(+)}|phys>=0.
\end{equation}
By the low-energy theorem the fermion pole term in $T_{\mu}^{n}$ is dominant
for the infrared singularity in $k_{n}^{\mu}$.These arise from diagrams in
which the soft-photon line with momentum $k_{n}^{\mu}$ and the incoming
fermion line can be separated from the remainder of the diagram by cutting a
single fermion line.Inclusion of regular terms and their contribution to
Ward-identities are given for the scalar case[1].Hereafter we consider the
one-photon matrix element $T_{1}$ which is given in [1,4,5]%
\begin{equation}
T_{1}=-ie\frac{1}{(r+k)\cdot\gamma-m+i\epsilon}\gamma_{\mu}\epsilon^{\mu
}(k,\lambda)\exp(i(r+k)\cdot x)U(r,s),
\end{equation}
where $U(r,s)$ is a four component free particle spinor with positive
energy.If we sum infinite numbers of photon in the final state as in
(5),assuming pole dominance for $k_{n}^{\mu}$ we have a simplest solution to
$T_{n}$ in (9)
\begin{equation}
T_{n}|_{k_{n}^{2}=0}=T_{1}T_{n-1}=e\frac{\gamma\cdot\epsilon_{n}}{\gamma
\cdot(r+k_{n})-m}T_{n-1}.\nonumber
\end{equation}
From this relation we obtain the $n$-photon matrix element $T_{n}$ as the
direct products of $T_{1}$%
\begin{equation}
\left\langle \Omega|\psi(x)|r;k_{1},...,k_{n}\right\rangle \left\langle
r;k_{1},..k_{n},\overline{\psi}(0)|\Omega\right\rangle \rightarrow%
{\displaystyle\prod\limits_{j=1}^{n}}
T_{1}(k_{j})\overline{T}_{1}(k_{j}).
\end{equation}
In this way we have an approximate solution of (5) by exponentiation of the
one-photon matrix element
\begin{equation}
\overline{\rho}(x)=-(i\gamma\cdot\partial+m)\int\frac{d^{2}r}{(2\pi)^{2}r^{0}%
}\exp(ir\cdot x)\exp(F),
\end{equation}%
\begin{equation}
F=\sum_{one\text{ }photon}\left\langle \Omega|\psi(x)|r;k\right\rangle
\left\langle r;k|\overline{\psi}(0)|\Omega\right\rangle =\int\frac{d^{3}%
k}{(2\pi)^{2}}\delta(k^{2})\theta(k^{0})\exp(ik\cdot x)\sum_{\lambda,s}%
T_{1}\overline{T_{1}}.\nonumber
\end{equation}
The fermion propagator is written explicitly in the following form in quenched
case%
\begin{align}
S_{F}(x)  &  =-(i\gamma\cdot\partial+m)\frac{\exp(-m\left\vert x\right\vert
)}{4\pi\left\vert x\right\vert }\nonumber\\
&  \times\exp(-e^{2}\int\frac{d^{3}k}{(2\pi)^{2}}\exp(ik\cdot x)\theta
(k^{0})\delta(k^{2})[\frac{m^{2}}{(r\cdot k)^{2}}+\frac{1}{(r\cdot k)}%
+\frac{d-1}{k^{2}}]),\nonumber\\
\left\vert x\right\vert  &  =\sqrt{-x^{2}}.
\end{align}

\subsection{\qquad\bigskip Explicit solution}

The function $F$ is evaluated with exponential cut-off for quenched case[1,3,4]%

\begin{equation}
F=\frac{e^{2}(d-2)}{8\pi\mu}+\frac{\gamma e^{2}}{8\pi m}+\frac{e^{2}}{8\pi
m}\ln(\mu\left\vert x\right\vert )-\frac{e^{2}}{8\pi}\left\vert x\right\vert
\ln(\mu\left\vert x\right\vert )-\frac{e^{2}}{16\pi}\left\vert x\right\vert
(d-3+2\gamma),
\end{equation}
where we find mass shift $\Delta m=\frac{e^{2}}{16\pi}(d-3+2\gamma
)+\frac{e^{2}}{8\pi}\ln(\mu\left\vert x\right\vert ).$Here we apply the
spectral function of dressed photon in the Landau gauge[2,6] to evaluate the
unquenched fermion proagator.
\begin{equation}
\rho^{D}(k)=\operatorname{Im}D_{F}(k)=\frac{c\sqrt{k^{2}}}{k^{2}(k^{2}+c^{2}%
)},c=\frac{e^{2}N}{8}.
\end{equation}
In this case wave function renormalization is given%
\begin{equation}
Z_{3}^{-1}=\int_{0}^{\infty}2\rho^{D}(\mu)\mu d\mu=\pi.
\end{equation}
We have an improved $F$
\begin{align}
\widetilde{F}  &  =\int_{0}^{\infty}2F(\mu)\rho^{D}(\mu)\mu d\mu=\frac{e^{2}%
}{8\pi c}(4)\ln(\frac{\mu}{c})+\frac{\gamma e^{2}}{8m}+\frac{e^{2}}{8m}%
\ln(c\left\vert x\right\vert )\nonumber\\
&  -\frac{e^{2}}{8}\left\vert x\right\vert \ln(c\left\vert x\right\vert
)-\frac{e^{2}}{16}\left\vert x\right\vert (3-2\gamma),
\end{align}
here linear divergent term is regularized by cut-off $\mu.$The fermion
propagator with $N$ flavours in position space is modified to
\begin{align}
S_{F}(x)  &  =-(i\gamma\cdot\partial+m)\overline{\rho}(x),\\
\overline{\rho}(x)  &  =\frac{\exp(-m_{0}\left\vert x\right\vert )}%
{4\pi\left\vert x\right\vert }\exp(\widetilde{F})=\frac{\exp(-\left\vert
m_{0}+B\right\vert \left\vert x\right\vert )}{4\pi\left\vert x\right\vert
}(c\left\vert x\right\vert )^{D-C\left\vert x\right\vert }(\frac{\mu}%
{c})^{\beta}\exp(\frac{\gamma e^{2}}{8m}),\nonumber
\end{align}%
\begin{equation}
B=\frac{c}{2N}(3-2\gamma),\beta=\frac{4}{N\pi},C=\frac{c}{N},D=\frac{c}%
{Nm},m=m_{0}+B.
\end{equation}

\subsection{ Momentum space}

It is known%
\begin{equation}
\int d^{3}x\exp(-ip\cdot x)\frac{\exp(-m\left\vert x\right\vert )}%
{4\pi\left\vert x\right\vert }(c\left\vert x\right\vert )^{D}=c^{D}%
\frac{\Gamma(D+1)\sin((D+1)\arctan(\sqrt{p^{2}}/m)}{\sqrt{p^{2}}(p^{2}%
+m^{2})^{(D+1)/2}}.\nonumber
\end{equation}
If we assume the spectral density for position dependent mass which makes
fermion fat%
\begin{equation}
\int_{c}^{\infty}\exp(-s\left\vert x\right\vert )M(s)ds=\left(  c\left\vert
x\right\vert \right)  ^{-C\left\vert x\right\vert },
\end{equation}
we get the propagator in Euclid space
\begin{equation}
S_{F}(p)=(\gamma\cdot p+m)Ac^{D}\Gamma(D+1)\int_{c}^{\infty}\frac
{M(s)\sin((D+1)\arctan(\sqrt{p^{2}}/(s+m))ds}{\sqrt{p^{2}}(p^{2}%
+(m+s)^{2})^{(D+1)/2}}.
\end{equation}%
\begin{equation}
S_{F}(p)=(\gamma\cdot p+m)Ac\int_{c}^{\infty}\frac{2M(s)(m+s)ds}%
{(p^{2}+(m+s)^{2})^{2}},(D=1).
\end{equation}

Explicit evaluation of $M(s)$ is not easy.In Fig.1 and Fig.2 we see the
profile of the scalar part of the propagator in position and momentum space
respectively for $D=1$ and quenched linear approximation to Dyson-Schwinger
equation[7].%
\begin{equation}
S_{F}(p)=\frac{\gamma\cdot p+m(p)}{p^{2}+m^{2}},m(p)=\frac{m^{3}}{p^{2}+m^{2}%
},m=\alpha=\frac{e^{2}}{4\pi}.
\end{equation}
Effects of position dependent mass with vacuum polarization correction can be
seen by comparison with the quenched linear approximation.%

{\parbox[b]{2.8513in}{\begin{center}
\includegraphics[
height=2.8513in,
width=2.8513in
]%
{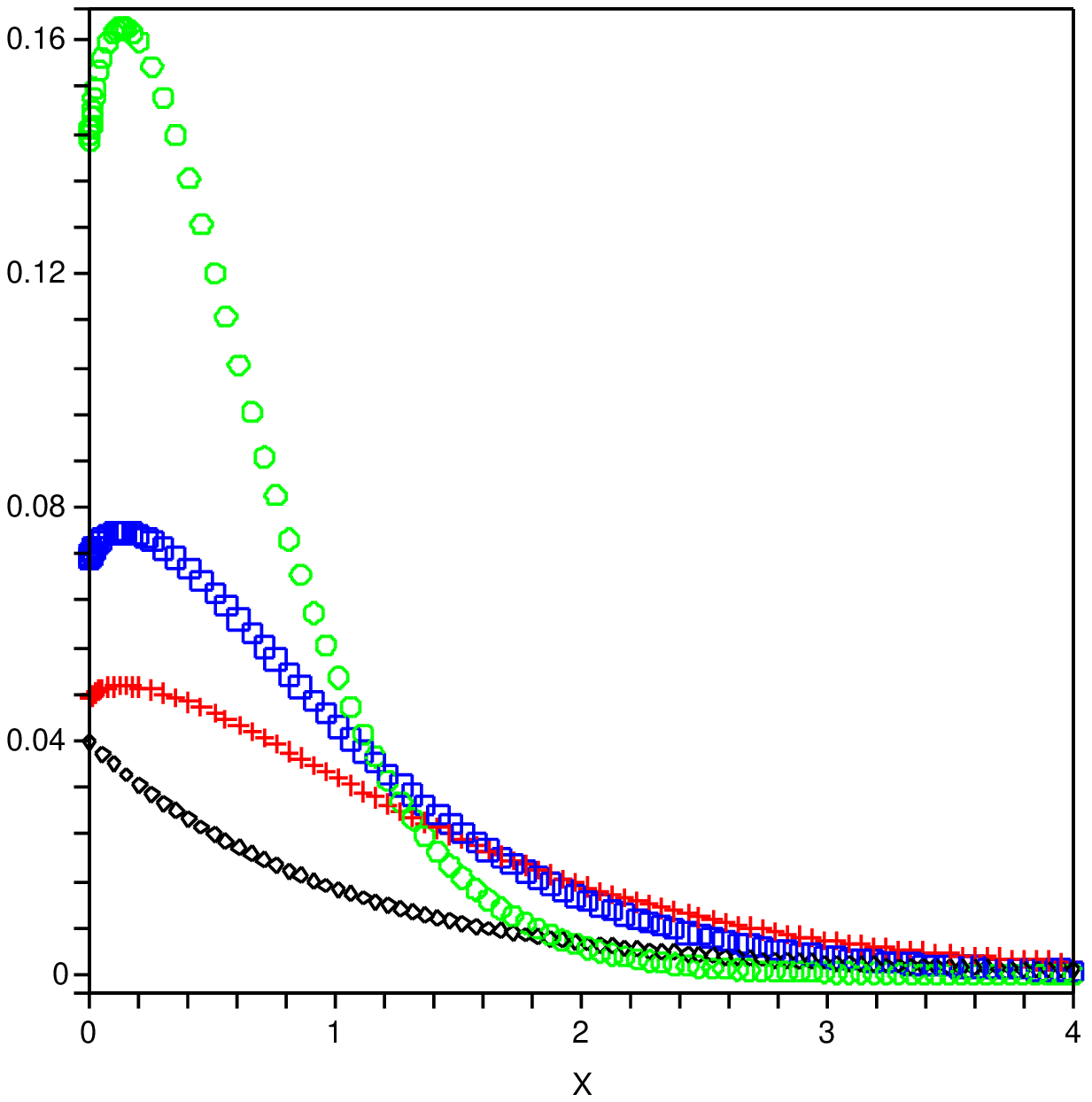}%
\\
Fig.1 $m\overline{\rho}(x)$ for $N=0(\circ),1(\bigcirc),2(\square),3(\times)$
in unit of $\alpha$ and $c.$%
\end{center}}}%
{\parbox[b]{2.8513in}{\begin{center}
\includegraphics[
height=2.8513in,
width=2.8513in
]%
{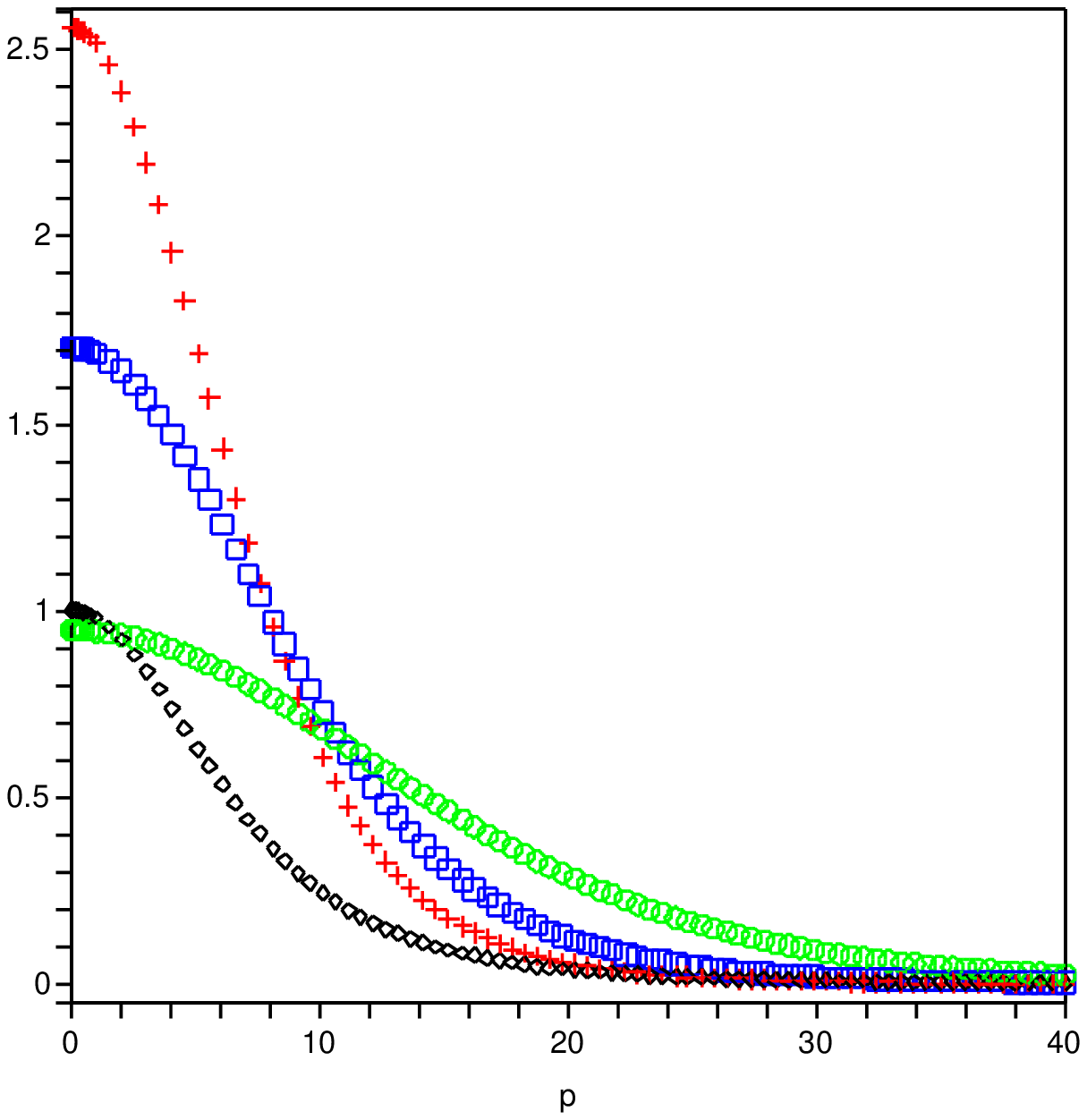}%
\\
Fig.2 $m\rho(p)$ for $N=0(\circ),1(\bigcirc),2(\square),3(\times)$ in unit of
$1/\alpha$ and $1/c,p$ in unit of $0.1.$%
\end{center}}}%

\section{\bigskip CONFINEMENT AND\ PAIR\ CONDENSATION}

We have the renormalization constant and bare mass in our approximation from
(22)
\begin{equation}
S_{F}^{0}=Z_{2}S_{F},\frac{Z_{2}^{-1}}{\gamma\cdot p-m_{0}},
\end{equation}%
\begin{equation}
Z_{2}^{-1}=\lim_{p\rightarrow\infty}\frac{1}{4}tr(\gamma\cdot pS_{F}%
(p))=0,m_{0}Z_{2}^{-1}=\lim_{p\rightarrow\infty}\frac{1}{4}tr(p^{2}%
S_{F}(p))=0.
\end{equation}
There is no pole and it shows the confinement for $D>0$.Order parameter
$\left\langle \overline{\psi}\psi\right\rangle $ is given in position space
directly from (20)%
\begin{align}
\left\langle \overline{\psi}\psi\right\rangle  &  =-trS_{F}(x)=4N\lim
_{x\rightarrow0}m\overline{\rho}(x)\nonumber\\
&  =-\frac{4Nmc^{D}}{4\pi}\exp(\frac{\gamma e^{2}}{8m})\lim_{x\rightarrow
0}x^{D-1}.
\end{align}
From the above equation we see that $\left\langle \overline{\psi}%
\psi\right\rangle $ $\neq0$ and finite only if $D=c/Nm=1.$

Here we consider the condition of $D=1$ is satisfied or not.In our
approximation there is not a definite way to determine physical mass $m.$The
linear approximation to the Dyson-Schwinger equation in quenched case leads
$D=1$[7].In our case for vanishing bare mass physical mass is assumed to be
\begin{equation}
m=\frac{c}{2N}(3-2\gamma),D=\frac{2}{3-2\gamma}=1.08.
\end{equation}
The condition is approximately satisfied.If we add small but finite bare mass
it may be satisfied.For $D=1,N=1$ case we have%
\begin{equation}
\left\langle \overline{\psi}\psi\right\rangle =-\frac{c^{2}}{\pi}\exp
(\gamma)=-8.885\times10^{-3}e^{4}%
\end{equation}
,which is the same order as in ref[8,9].

\section{SUMMARY}

We evaluated the non-perturabative fermion propagator in QED$_{3}$ by the
method of spectral function.Infrared behaviour was significantly modified by
position dependent mass for $N=1$ case in comparison with quenched linear
aaproximation to Dyson-Scwinger case.It is very slow to reach high energy
behaviour $1/p^{4}.$At large $N$ these effects are suppresed and damps fast as
$1/p^{4}.$

\section{\bigskip REFERENCE}

[1]R.Jackiw,L.Soloviev,Phys.Rev.\textbf{137}.3(1968)1485.

[2]A.B.Waites,R.Delbourgo,\ Int.J.Mod.Phys.\textbf{A7}(1992)6857.\newline

[3]L.S.Brown,Quantum field theory,Cambridege University Press,1992.\newline

[4]Y.Hoshino,$Nucl.Phys${\normalsize .}\textbf{B(}Proc.Supple.\textbf{)161(}%
2006\textbf{)}95-101.\newline

[5]K.Nishijima,$Fields$ $and$ $Particles$,W.A.Benjamin,Inc.,1969.\newline

[6]C.Itzykson,J.B.Zuber,Quantum field theory,McGRAW-HILL,1980.

[7]M.Koopman,Dynamical mass generation in QED$_{3}$,Ph.D thesis,GroningenUniversity(1990),

\ \ \ chapter 4.

[8]Y.Hoshino,T.Matsuyama,C.Habe,$Fermion$ $Mass$ $Generation$ $in$ $QED_{3}$,

\ \ \ in the Proceedings of the Workshop on Dynamical Symmetry Breaking,

\ \ \ Nagoya(1989)0227-232.

[9]C.S.Fischer,R.Alkofer,T.Dahm,and P.Maris,Phys.Rev.\textbf{D70(}%
2004\textbf{)}

\ \ \ 073007 [arXiv:hep-th/0407104].\newline

\end{document}